\documentstyle[12pt]{article}

\author{Alfred H\"ubler\thanks{Also at Center for Complex Systems Research,
Beckman Institute, UIUC.}  and David Pines\\
The Santa Fe Institute,\\
1660 Old Pecos Trail, Santa Fe, NM 87501\\
and\\
Physics Department, University of Illinois at
Urbana-Champaign\thanks{Permanent address.}\\
1110 West Green Street, Urbana, IL 61801-3080\\}

\title{Prediction and Adaptation in an Evolving Chaotic Environment}

\begin{document}
\maketitle
\newpage
\begin{abstract}
We describe the results of analytic calculations and computer
simulations of adaptive predictors (predictive agents) responding to
an evolving chaotic environment and to one another.  Our simulations
are designed to quantify adaptation and to explore co-adaptation for a
simple calculable model of a complex adaptive system.  We first
consider the ability of a single agent, exposed to a chaotic
environment, to model, control, and predict the future states of that
environment.  We then introduce a second agent which, in attempting to
model and control both the chaotic environment and the first agent,
modifies the extent to which that agent can identify patterns and
exercise control. We find that (i) optimal adaptive predictors have an
optimal memory and an optimal complexity, which are small for a a
rapidly changing map dynamics and (ii) that the predictive power can
be increased by imposing chaos or random noise onto the map dynamics.
The competition between the two predictive agents can lead either to
chaos, or to metastable emergent behavior, best described as a
leader-follower relationship.  Our results suggest a correlation
between optimal adaptation, optimal complexity, and emergent behavior,
and provide preliminary support for the concept of optimal
co-adaptation near the edge of chaos.

\end{abstract}
\newpage
\section{Introduction}
In recent years various examples of complex adaptive systems, such as neural nets \cite{pruning}, genetic
algorithms\cite{holland,rechenberg}, and evolutionary models including
artificial life \cite{langton} have been used in carrying out computational optimizations. In addition adaptive
algorithms and evolutionary models have been used to mimic the
evolution of biological and social systems
\cite{mittenthal,arrow,brock}. When used for the prediction of a
complex dynamics common features, such as a tendency to evolve their internal dynamics to the edge of chaos
\cite{edgelangton,edgepackard,edgekauffman,crutchfield} and the emergence of adaptive predictors  with a bounded
rationality or a limited memory \cite{arthur} have been observed in many of
those systems.
However it is
typical for the systems which have been studied that they have many parameters and often
their properties depend sensitively on the adjustment of those
parameters \cite{wolfram}.

As a first step toward the development of an approach which
incorporates a quantitative understanding of common features of complex
adaptive systems, we have
carried out computer simulations of a simple "toy model" of a complex
adaptive system: individual agents, operating in an evolving chaotic
environment specified by a logistic map, seek to predict the future
states of their environment by modeling and controlling it. 
We study the dynamics of adaptive
predictive agents responding to an evolving chaotic environment and to one
another.  Our simulations are designed to quantify adaptation and to
explore co-adaptation for a simple calculable model of a complex
adaptive system.  

Although elementary, and
in large part calculable, our adaptive agents
(predictors) meet
the definition of a complex adaptive system proposed by Murray
Gell-Mann\cite{gellman}:
\begin{itemize}
\item information gathering entities 
\item respond both to the environment and to one another
\item 	segregate information from random noise
\item 	compress regularities into a model (schema)
\item 	modify their internal characteristics to
improve
their predictive (adaptive) capacity
\end{itemize}

In our model the chaotic environment to which an agent responds is
specified by a simple logistic map, with parameters which can be
altered, plus random or dynamic noise which can also be
altered.  A given agent may be either passive or active; thus
agents both respond to the environment (by receiving signals from it)
and attempt to control it (by sending signals to
it)\cite{hublerluscher}.  More specifically, an agent measures
and models the chaotic environment and
employs
various control strategies to predict its future states. 
For each agent we give explicit quantitative measures of:
\begin{itemize}
\item 	adaptation (the predictive ability of the agent)
\item	complexity (the number of parameters used to specify an agent's
model)
\item 	memory (the data used in the modelling process)
\end{itemize}
We determine both experimentally (via our computer simulations) and
analytically the conditions for optimal predictive behavior
(adaptation),
complexity, memory.

We first consider the ability of a single agent,
exposed to a chaotic environment, to model, control, and predict the future
states of that
environment.  We then introduce a second agent which, in attempting to
model and control both the chaotic environment and the first agent,
modifies the extent to which that agent can identify patterns and
exercise control.  

Our computer simulations demonstrate the consequences of
competition between the two agents. Competition leads to
chaos, if the agents follow typical learning strategies, or to
emergent metastable behavior, if the agents develop a
new learning strategy.  Thus, we find metastable solutions (strategies)
in which the two agents optimize their joint predictive capacities
by co-adapting in a leader-follower relationship.  A sufficient
condition for arriving at this joint strategy is the development of
adaptive predictions which enable one agent to recognize the presence
of another.  Our results suggest a correlation between optimal
adaptation, optimal complexity, and emergent behavior. 
Preliminary support is provided for the concept of optimal co-adaptation
near an
order-disorder
transition\cite{edgelangton,edgepackard,edgekauffman,crutchfield}

The computer simulations were performed
on a Silicon Graphics 340 VGX machine.  The
length of time for a given study varies from
1000 time steps (the number required to determine numerically the
optimal response of a single agent) to 100,000 time steps (the number
required to explore in detail the competition between two active
agents which leads, over time, to their arriving at dynamic controls
near the edge of chaos).  For two agents, each time step required 
$2$ sec of cpu time.

In Sec. 2 we specify our model, and consider the behavior of a single
agent.  We consider competition (and cooperation) between two agents
in Sec~3. Extensions and possible applications are discussed in Sec 4.

\section{Systems with One Agent}

\subsection{The Dynamics of the Environment}

We consider an environment described by a simple logistic map.The state of the environment at some time $n$ determines
its state at a later time $n+1$:
\begin{equation}
y_{n+1}=p y_{n}(1-y_{n}),
\end{equation}
where $0\le y_{n}\le 1$ represents the state of the environment at
time $n$ , and $1<p<4$ is a control parameter. Each transition from
$y_n$ to $y_{n+1}$ is an event.  Depending on the choice of
$p$, the environment may be stationary, periodic, or chaotic in nature.
The dynamics of the logistic map converges to a non zero
fixed point for $1\le p\le 3$,labelled as stationary in our plots; it
converges to a period 2-cycle for $3 \le p \le 3.449449\ldots$, labelled
as periodic, and to more and more complicated period-$2^{\beta}$
cycles, $\beta=2,3,\ldots$, for $3.449 449 \ldots \le p \le p^{crit}$.
Above $p^{crit} = 3.569 946\ldots$ the attractor of the map is
chaotic, except in special ``periodic windows''\cite{eajackson}. In
our plots this last parameter region is labelled as chaotic, since
most of our simulations are done at a large noise level, where the
periodic windows have little impact on the dynamics \cite{gmk}. The Liapunov
exponent of the map, defined by
\begin{equation}
\lambda^{e}
=\lim_{N\rightarrow\infty}\frac{1}{N}\sum_{n=1}^{\infty} \ln |\frac{d
f(y)}{dy}\vert_{y=y_n}|,
\end{equation}
is positive for
$p_{1,n}>p_{1,n}^{critical}$ (apart from periodic windows), and is negative otherwise; here $f(y)=p
y (1-y)$.

The use of a simple logistic map to describe a typical
high dimensional complex environment may be considered an
oversimplification. However, there is a long history of
similar approaches in studies of physical systems.  Most
physical systems are in reality high dimensional systems. A
physical pendulum, often used as an example of low
dimensional motion, has many degrees of freedom. When
stimulated by a short kick, such as the impact of an hammer, many
kinds of vibrations may be stimulated.  Usually those vibrations die
out fast, which means that the dynamics settles down to a low
dimensional approximate inertial manifold. 
Often, the trajectories are complicated but
confined to a
very small region on this approximate inertial manifold.  If the
inertial manifold and the flow vector field are smooth in this region,
one may expand the flow vector field in a Taylor series and drop
higher order terms.  In this case, the limiting dynamics is low
dimensional and the nonlinearity of the corresponding flow vector
field is of low order. Lorenz has shown\cite{lorenz} that low
dimensional, low order systems can exhibit deterministic chaos, {\em
i.e.} irregular motion which is sensitive to initial conditions.
Since the flow vector field of such systems is smooth, their dynamics
is usually smooth and oscillatory with trajectories which may have a
simple or fractal
geometry\cite{feigenbaum,grassberger}. Moreover, Poincare
\cite{poincare} has shown that it is in general useful to study the
dynamics of the amplitudes of the smooth oscillatory 
motion of a low dimensional, low order system and to model it with low
order maps. Therefore, we model the environment with
logistic map dynamics, a simple, nontrivial
deterministic chaotic system.

In many systems of interest, noise is present and the control
parameters vary over time. We thus wish to consider
environments in which additive background noise is present, and in
which the control parameter is noisy:
\begin{eqnarray}
y_{n+1}=p_n y_n (1-y_n)+F_{n}^{\sigma_{D}}\label{envnoise1}\\
p_{n+1}=p_{n} + F_{n}^{\sigma_p},
\end{eqnarray}
where $F_{n}^{\sigma_D}$ describes additive system noise and,
$F_{n}^{\sigma_p}$ is parametric noise.  The noise parameters
$F_{n}^{\sigma_D}$, $F_{n}^{\sigma_p}$ have a mean which is zero and
standard deviations, $\sigma_{D}$ and $\sigma_p$. $\sigma_{p}$
determines the rate of change of the environment, whereas $\sigma_D$
measures the noise level. We assume that the rate of change of the control
parameter of the
environment is small, {\em i.e.} $\sigma_{p} \ll 1$. Fig.
\ref{fignoise} illustrates the dynamics of such an environment.

\subsection{The Dynamics of the Agent}

\subsubsection{Prediction, Adaptation, Learning, and Innovation}

In our numerical experiments, the primary goal of each agent is
prediction of the environment to which it and, where present,
other possible agents are responding. An agent seeks to discover
patterns or regularities in the environment; this process of
constructing a model of the environment is greatly facilitated if the
agent has the option of turning on or off a control signal which may
entrain the environmental dynamics to a predetermined sequence of
states, defined as the {\bf goal dynamics} of the agent, a
process of active adaptation.  As we shall see,
by pursuing a reasonably well defined strategy, a single agent can  
arrive at a goal dynamics which maximizes its predictive power.

A given agent samples a set of
$N_n$ successive events $y_i$, $y_{i+1},\ldots$,$y_{i+N_n}$, which
characterize the time evolution of the environment and uses it to
predict the next $m$ events, $\tilde{y}_{i+N_n+1}$, $\tilde{y}_{i+N_n+2},
\ldots$, $\tilde{y}_{i+N_n+m}$
The success of the prediction process of the agent is
measured by the {\bf prediction error}, $\epsilon_{n,m,M}$ defined
with respect to the background noise level, $\sigma_D$, as:
\begin{equation}\label{enmm}
\epsilon_{n,m,M} = \frac{1}{M} \sum_{i=n}^{n+M}
(\tilde{y}_{i+m}(y_i) - y_{i+m})^2/ \sigma_D^2,
\end{equation}
where $\tilde{y}_{i+m}$ is a $m$-step prediction of the environment by
the agent. The prediction error is a sliding average of length
$M$ over an ensemble of rapidly fluctuating values. It depends on the
time of prediction, $n$, and the number of steps predicted, $m$.
The best prediction is
limited by the background noise level $\sigma_D$;. From
Eq.~\ref{enmm} the minimum value of the prediction
error is $1$.

At each step $n$, the agent can modify its model. This process of
modification, or updating, represents the {\bf adaptive behavior} of
the agent. For our adaptive agents, {\bf learning} is a two step process:
\begin{itemize}
\item first, acquire and apply a predetermined class
of schemata or fitting functions to model the past events of the
environment 
\item second, choose
fitting function parameters that do the best job of prediction.
\end{itemize}

In the present context, {\bf innovation} involves the development
and application of a new and different class of models to analyze the
past and predict the future. It may include active
adaptation, an exploration of the response of the environment to an
imposed goal dynamics.  We assume that the time scale of the
innovation process is in general longer than the time scale of the
learning process $N_n$. We also assume that the probability an agent
decides to try to improve a parameter of the modelling and control
processes is given by:
\begin{eqnarray}\label{P}
P_{{n}} = c \epsilon_{n,m,M}.
\end{eqnarray}
\noindent Here $c$ (a constant) is the minimum rate
of adaptation and $\epsilon_{n,m,M}$ is the current $m$ step
prediction error of the agent averaged over $M$ time steps. The
minimum rate of adaptation, $c$, may be
different for each parameter of the modelling and control processes.
Typical values of $c$ are such that $10^{-6}\le c \le 10^{-3}$.

The agent begins with a given initial setting of the model and
control processes and observes the environment for at least $N_n$ time
steps. It then calculates an initial model and updates the model and
prediction error at each subsequent time step. In addition, at each
time step, the agent selects a random number, $R$, which lies between
$0$ and $1$. If $P_n< R$, the agent does nothing; if $P_n \ge R$ the
agent innovates. With this procedure, the probability that an agent
will innovate is simply $P_n$, as long as $P_n\le 1$. Whenever $P_n>1$, the
agent will always innovate. Thus after, say, $t$
time steps the agent will innovate, {\em i.e.} try another class of
fitting functions, switch the control on, or alter the goal
dynamics. $t$ is short if the prediction error is large and vice versa
according to Eq.\ref{P}. The expectation value of $t$ is $P_n^{-1}$.
Eq.\ref{P} also guarantees perpetual novelty;
no matter how well an agent is doing, that agent will,
sooner or later, be prompted to innovate.
For
optimally predictive agents, whose prediction error is of order unity, the
minimum
rate of innovation is $c$.  The trial period for innovation is
assumed to last $2M$ steps. After that period the parameter is
reset to its previous value if the prediction error has increased
on average during the last $M$ steps of the trial period.

\subsubsection{Learning: Modeling and Prediction}

An agent develops a model, $f^{model}_{{n}}$, of the environmental
dynamics
for single and multiple step predictions of the environment
\begin{equation}
\tilde{y}_{n,m}(\tilde{y}_{n,m-1})=
f^{model}_{{n}}(\tilde{y}_{n,m-1})
\end{equation}
where $\tilde{y}_{n,0} = y_n$ and where $m = 1,2,3\ldots$ counts the
number of steps. This is based on the assumption that a good
predictor for the observed events is a good predictor for future events.

The first step of the modelling process is for the agent to represent
the observed events in a state space (see
Fig.\ref{interpolation}). In
the simulations, we restrict the attention of each agent
to events that lie within a region of
interest, a range of events such that
$y_{{n}}^{min,I}\le y \le y_{{n}}^{max,I}$. $y_{{n}}^{min,I}$ and
$y_{{n}}^{max,I}$ are the boundaries of the region of interest at time
step $n$.  In {\bf modeling} the environmental dynamics each agent
uses those $N_{{n}}$ events
which are most recent and in which the initial state is in the region
of interest.  In our numerical examples the region of interest is
usually slightly larger than the region where events have been
observed during the first 1000 time steps of each simulation. Further
we introduce a 
equidistant grid $y^{grid}_{j,{n}}$, j=1,..,$N_{G}$, $N_{G} \gg
N_{n_m}$ and estimate the events $(y_{j,{n}}^{grid},f_{j,{n}}^{grid})$
at these grid points through linear interpolation\cite{adaptation} (see Fig.
\ref{interpolation}).  The interpolation represents a generalization
of the observed events, since the agent is guessing the behavior of
the environment for those states where no observations are available.

The relation between $y_{j,{n}}^{grid}$ and
$f_{j,{n}}^{grid}$ is represented by a Fourier series:
\begin{eqnarray}\label{model}
f^{model}_{{n}}(y_{n}) & \equiv & \sum^{N_G}_{k = 1}p_{k,{n}}^{model}
\sin \left(\frac{\pi}
{y_{{n}}^{max,I}-y_{{n}}^{min,I}} k (y_{n}-y_{{n}}^{min,I})\right)
\nonumber\\
& & \mbox{} + d_{0,{n}}^{model} + d_{1,{n}}^{model} (y_{n}-y_{{n}}^{min,I})
\end{eqnarray}
where $d_{0,{n}}^{model}$ and $d_{1,{n}}^{model}$ are parameters of
the Fourier analysis chosen to improve the convergence of the
Fourier series.  Unless we specify otherwise, we assume that the
observed events are almost homogeneously distributed in the region
of interest, and that interpolation errors are small compared to
statistical errors. In this case the standard deviation
$\sigma_{p_{k}}$ of the Fourier coefficients is given by
$\sigma_{p_k}=\sigma_D\sqrt{\frac{2}{N_n}}$.

The last step of the modeling process is to compress the information
which is contained in the generalized observed events by segregating
information from random noise. The values of the model parameters
$p_{k,{n}}^{model}$ are determined by a least-squares fit which
minimizes the prediction error by minimizing the difference between
the generalized events at zero noise level and a model for single step
predictions (Eq. \ref{model}) for the observed events.  If we
assume that the observed states are homogeneously distributed in the
region of interest, that the additive noise is uncorrelated and that
the rate of change of the environment is small, the fit problem has a
unique solution and the optimal parameters $p_{k,{n}}^{model}$ are
given given by the Fourier coefficients $\tilde{p}_{k,n}^{model}$ of
$f_{j,{n}}^{grid}$\cite{adaptation}, or are equal to zero, if the Fourier
coefficient is smaller than its standard deviation $\sigma_{p_k}$.  An
analogous procedure, pruning, is followed in neural nets, where
it is found to improve their performance\cite{pruning}. If we define {\bf complexity}, $K_n$, as the number of model parameters
used by the agent, it is possible to determine the complexity
$K_n^{opt}$ of an optimal model. The optimal model neglects all parameters
with value smaller than the error bar for that parameter, estimated
by its standard deviation. This concept is illustrated in
Fig.~\ref{optimal}$a,b$.  The minimum prediction error in
Fig.~\ref{optimal}$a$ is at approximately $k=10$.  In Fig.~\ref{optimal}$b$
the Fourier coefficients equal
the error bar also at $k=10$.  The optimal complexity is small for a
large rate of change of the environment and vice versa.

The environmental dynamics is a parabolic function with only one parameter.
Therefore, a Tschebycheff series or a Legendre series would
converge even faster than the Fourier series since these are also
polynomial. However we intentionally program the agents to use a
Fourier series in order to illustrate the point that an exact match
between the set of models (schemata) and environmental dynamics is not
necessary.

The Fourier coefficients of many continuous, piecewise linear
functions converge parabolically\cite{bs}, {\em i.e.}
$\tilde{p}_{k,n}^{model} \approx \tilde{p}_{c,n}^{model}/k^2$, with an
appropriate choice of $d_{0,{n}}^{model}$ and $d_{1,{n}}^{model}$ and
converge linearly otherwise {\em i.e.} $\tilde{p}_{k,n}^{model}
\approx \tilde{p}_{c,n}^{model}/k$, where $\tilde{p}_{c,n}^{model}$ is
a number.  
For parabolic convergence an
estimate of $K_n^{opt}$ is given by
\begin{equation}
K_n^{opt} = (
\frac{\tilde{p}_{c,n}^{model}\sqrt{N_n}}{\sqrt{2}\sigma_D})^{1/2}
\end{equation}

The prediction error for an agent with complexity $K_n$ can be estimated by
writing
\begin{eqnarray}\label{eqeps}
&&\epsilon_{n,1} = 1 + K_n \frac{1}{N_n} + \sum_k
\frac{(\tilde{p}_{k,n}^{model})^2}{2\sigma_D^2}\delta_{k,n} 
+ \sum_k (\frac{p_{k,n}^{model}\sigma_p}{p_n\sigma_D})^2
\frac{N_n}{6}\label{e1}\\
&&\epsilon_{n,m+1} = \epsilon_{n,m} \exp{(2\lambda^{e})}
\end{eqnarray}
where $m=1,2,\ldots$ and where $\delta_{k,n} = 1$ if
$p_{k,{n}}^{model}=0$ and $\delta_{k,n} = 0$ otherwise. 

Another quantity which can be optimized by an agent is the number
$N_n$ of states of the environment which are used in the
modelling process.  In principle, an agent is assumed to
have access to the whole history of the environmental dynamics. In
practice, the agent will find it advantageous to use only a small
portion of this information.  Since $N_n$ measures how much of this
information is used for the modeling process, $N_n$ is a measure for
the {\bf memory} of the agent.  If $N_n$
is large, outdated data may decrease the quality of the
model. If $N_n$ is too small, statistical
errors may prevent an agent from choosing an optimal description.
Therefore, there is an {\bf optimal memory} $N_n^{opt}$ of an
agent, which strikes a balance between the errors
introduced by the noise level and those produced by the rate of change
of the environment. Fig. \ref{optimal}$c$ shows how a proper choice of
$N_n$ minimizes the prediction error.

The last term of Eq.\ref{e1} increases with the rate of change of the
environment $\sigma_p$ and $N_n$.
Since the first term in Eq.
\ref{e1} decreases with $N_n$, the prediction error $\epsilon_{n,l}$
is minimal for
\begin{equation}
N_n = N_n^{opt} = (\frac{6K_n^{opt}}{\sum_k
(\frac{p_{k,n}^{model}}{p_n})^2})^{\frac{1}{2}} \frac{\sigma_D}{\sigma_p}
\end{equation}
\ \\
as depicted in Fig. \ref{optimal}. Agents
with optimal memory and optimal complexity possess a prediction error,
\begin{equation}\label{eqeps2}
\epsilon_{n,1}^{opt} = 1 + 2 (\frac{K_n^{opt}\sum_k
(\frac{p_{k,n}^{model}}{p_n})^2}{6})^{1/2} \frac{\sigma_p}{\sigma_D}
+ \frac{(\tilde{p}_{k,n}^{model})^2}{2\sigma_D^2}\delta_{k,n} 
\end{equation}

The noise level in the environment and the rate of change of
the environment determine the optimal memory of the agent. If the
Fourier series converges parabolically the optimal complexity of the
agent with optimal memory is:

\begin{equation}\label{kopt}
K_n^{opt} =
(\frac{3(\tilde{p}_{c,n}^{model})^4}{2\sum_k
(\frac{p_{k,n}^{model}}{p_n})^2\sigma_D^2\sigma_p^2})^{1/7}
\end{equation}

Since $N_n$ data are used to fit the model, a delay of $N_n$ time steps is
needed to model the environment after a sudden change
of the parameter $p_n$. Therefore, it is possible to establish a
relation between the complexity of the agent's model and the {\bf
learning rate}, the minimum time required to extract a
completely new model from the environmental dynamics. This
result provides a method to determine experimentally
whether an adaptive agent is functioning optimally. To evaluate an
agent's performance
an observer can introduce a sudden change of the environment and
then measure the recovery of an adaptive agent, as shown in
Fig.\ref{recovery}. A match
between the recovery rate and the optimal learning rate
indicates that the agent is optimally adapted.

An important feature of the modeling process is that the resulting
model parameters are unique. If the relation between the evolving
control parameters of the environmental dynamics and the model
parameters is continuous, it may be possible to apply the same
modeling procedure to the time series of the model parameters and to
construct hierarchical models, of the kind considered by
Crutchfield\cite{crutchfield}.

\subsubsection{Control of the Environment}

The process just described represents passive adaptive behavior
of the agent. However, an agent can modify the
environment in an effort to improve his predictive power, by turning
on or off a control signal, which may entrain the environmental
dynamics to a particular goal dynamics. Our control strategy for
such an active agent is based on the approach developed by one of us
\cite{alfredphd} for the control of chaos. The agent applies a
suitably chosen driving force to {\bf entrain} the chaotic environment
to a predetermined goal dynamics.

There is a close relation between entrainment and optimal prediction.
Entrained oscillators may have an optimal energy exchange since they
are at resonance \cite{eisenhammer} while the concept of optimal
information
transfer has widespread application in research on phase locked
loops\cite{lichtenberg,chua}. As we shall see, for the problem at hand
entrainment makes optimal prediction possible.

To control the environmental dynamics an active adaptive agent
iterates a logistic map time series $x_n$ of desired environmental
states; this {\bf goal dynamics} is specified by a parameter $p_n^{goal}$:

\begin{equation}
x_{n+1} = g(x_{n},p_{n}^{goal})
\end{equation}
\\
to impose the goal dynamics on the environment, the agent applies a driving
force; 
\begin{equation}\label{force}
F_{n} = x_{n+1} - f_{{n}}^{model}(x_{n}),
\end{equation}
tailored to make up the
difference between the agent's model of the uncontrolled environmental
dynamics and the desired state of the environment. The resulting controlled
environment dynamics is given by:
\begin{eqnarray}\label{envcontrol}
&&y_{n+1}=p_n y_n (1-y_n)+F_{n}+F_{n}^{\sigma_{D}}\\
&&p_{n}=p_{n} + F_{n}^{\sigma_p},\label{envcontrol3}
\end{eqnarray}

Control is advantageous. It enables the agent to avoid the
exponential growth of the prediction error with the number of steps,
found in the case of chaotic systems with positive Liapunov exponents
(see Fig. \ref{sp} and Fig.\ref{sc}).  If the agent succeeds in
controlling the environment, the prediction error becomes bounded; the
upper boundary becomes small if the control which is exercised is
stable over long periods of time. For an
example of the way in which control reduces the prediction error, see
Fig.\ref{control}. 

The type of the goal dynamics has both a direct and indirect impact on the
prediction error. Since the size of prediction error depends on the
stability of the control, an agent may improve the prediction error by
choosing a goal dynamics which provides a very stable control. From
this point of view, the prediction error would be as small as possible
if the goal dynamics is, or is close to a stationary state, or some other
superstable stationary orbit\cite{schuster} of
the unperturbed system.

However, this discussion takes only statistical errors into account.
If the goal dynamics is a stationary state, for example at $x_n=0.5$,
the interpolation procedure may lead to large systematic errors. This
is illustrated in Fig. \ref{withcontrol}. There we assume the region
of interest is the whole interval and that $f$ is known at the
boundaries of the interval.  Fig. \ref{withcontrol}a shows that the
prediction error may be significantly higher than the statistical
estimate as long as the goal dynamics is not chaotic, {\em i.e.} $p_n
\le 3.56$. This is because the linear interpolation
produces edges which are very sharp for stationary states close to
$x_n=0.5$ (Fig. \ref{withcontrol}b), much less sharp for period two
dynamics, and essentially absent for a chaotic goal dynamics. Of
course, other interpolation schemes could weaken this effect, but the
best solution to this problem is to pick a chaotic or random goal
dynamics which covers the entire state space and makes interpolations
unnecessary. Likewise, a small amount of additive noise in the
environmental dynamics or the goal dynamics may help to reduce the
prediction error, since it reduces systematic errors in the modelling
process\cite{breeden}. Moreover a control with a chaotic goal dynamics
makes the system more robust against sudden changes in the noise
level, since the agent has a global model of the flow vector field.

\section{Systems with two Agents}

As might be expected, the results change dramatically when two agents are
present.  For example, a second agent may alter the environmental
dynamics of the first agent sufficiently to make it impossible for the
latter to exercise effective control and make accurate predictions.
Or, without establishing direct communication, one agent may identify
the presence of a second, and the two may 
establish a cooperative relationship which improves their joint
predictive abilities.

We assume that the agents have the same region of interest but may
have different goal dynamics, and different models of the
environmental dynamics. To distinguish between the parameters of the
two agents, we attach a superscript $a=1,2$ to the model parameters
$p_n^{model,a}$, the memory $N_n^a$, the complexity, $K_n^a$, the
parameter of the goal dynamics $p_n^a$, the ON/OFF switch of the
control, $C_n^a$, the Liapunov exponent of the goal dynamics
$\lambda_n^{g,a}$, the control coefficient $\lambda_n^{c,a}$, the
prediction error $\epsilon_{n,m}^a$ and the control force $F_N^a$. The
dynamics of the environmental system is then given by:

\begin{eqnarray}\label{envcontrol2}
&&y_{n+1}=p_n y_n (1-y_n)+F_{n}^1+F_{n}^2+F_{n}^{\sigma_{D}}\\
&&p_{n}=p_{n} + F_{n}^{\sigma_p},
\end{eqnarray}
In the following we discuss the prediction errors of the two agents
for the situation where both are passive, {\em i.e.} both have their
control switched off, both are active, {\em i.e.} both have their
control switched on, and the leader follower situation, in which one
is active (leader) and one is passive(follower).
We study the lifetime of those structures, {\em i.e.} the
number of time steps between the start and end of such a configuration,
where trial periods do not count as the end of a configuration if this
configuration reemerges after the trial period. Unless we specify
otherwise, we assume that both agents have a memory and complexity
which would be optimal for single agent systems.

There are three scenarios for dual agent behavior: both may be
passive; one may be passive while the other is active, i.e.  exercises
control to improve its predictive power, a leader-follower situation;
or both may be active, vying for control (and optimal predictive
power).  We consider these scenarios in turn, assuming that each agent
is capable of optimizing its memory and complexity in response to
an environment in which the other agent is absent.

\subsection{Two Passive Agents}

When both agents are passive, the prediction error of each will
be as though the other agent were not present.  
 For a chaotic
environmental dynamics the prediction error is usually very large
compared to the leader follower situation. Following a trial
period, one or the other agent will turn on its control, and the
scenario becomes one of leader (the active agent) and follower
(the passive agent).

The life time of a configuration with two passive agents may be
estimated by:
\begin{eqnarray}\label{lpassiv}
L_{{n,m}} = \frac{c^2}{\epsilon_{n,m}^1 \epsilon_{n,m}^2}
\end{eqnarray}
where the prediction errors are the same as of single passive agents.
For a chaotic environment, passive agents have a large prediction
error. Therefore  the lifetime of a configuration with two passive
agents is quite short.

\subsection{The Leader - Follower Relation}

As long as the second agent remains passive, the scenario for the
behavior of the first agent, the leader, is identical to
that of a single agent.  The leader explores various
controls, entrains the environment, and improves his predictive power.
However, as this process of active adaptation proceeds, the
passive second agent, the follower, senses a changed environment.  
If the leader switches off its
control infrequently, the follower will
see an environment which is mainly determined by the goal dynamics of
the control exercised by the leader. Such a controlled environment is
easier to predict as long as the goal dynamics of the leader is
not chaotic.  Under these
circumstances, the optimal memory of the follower may, in fact,
exceed that of the leader, whose memory depends on both the noise
level and rate of change of the environment.  The results of our numerical
experiments on the role played by the goal dynamics of the leader are
displayed in Fig.\ref{predlf}.

In the {\bf leader-follower situation}, where one agent is active and
one agent is passive, the prediction error can be estimated by:
\begin{eqnarray}
&&\epsilon_{n,1}^1 = 1 + \exp{(2\lambda^{c,1})}  +
K_n \frac{1}{N_n} + \sum_k
\frac{(\tilde{p}_{k,n}^{model,1})^2}{2\sigma_D^2}\delta_{k,n}\nonumber\\
&&\mbox{\hphantom{$\epsilon_{n,1}^1 =$ }}
+\sum_k (\frac{p_{k,n}^{model,1}\sigma_p}{p_n}\sigma_D)^2
\frac{N_n}{6}\label{epscc1}\\
&&\epsilon_{n,m+1}^1 = \epsilon_{n,m}^1 \exp{(2\lambda^{c,1})} +
\epsilon_{n,1}^1\label{epscc2b}
\end{eqnarray}
for the leader, and by
\begin{eqnarray}
&&\epsilon_{n,1}^2 = 1 + 
K_n \frac{1}{N_n} + \sum_k
\frac{(\tilde{p}_{k,n}^{model,2})^2}{2\sigma_D^2}\delta_{k,n}  +
\sum_k (\frac{p_{k,n}^{model,2}\sigma_p}{p_n}\sigma_D)^2
\frac{N_n}{6}\label{epscc1b}\\
&&\epsilon_{n,m+1}^2 = \epsilon_{n,m}^2 \exp{(2\lambda^{g,1})} +
\epsilon_{n,1}^2\label{epscc2c}
\end{eqnarray}
for the follower. The lifetime 
of the leader follower configuration can be estimated by:
\begin{equation}\label{llf}
\renewcommand{\arraystretch}{2}
L_{{n,m}}  =  \left\{ \begin{array}{lll}
 \displaystyle\frac{c}{\epsilon_{n,m}^2} \mbox{ } \frac{p^{max}-p^{crit} +
p_n^1}{p^{max}
- p^{min}} & \mbox{ for} &  p_n^1 < p^{crit}\\
 \displaystyle\frac{c}{\epsilon_{n,m}^2} & \mbox{ otherwise }\\
                         \end{array} \right.
\end{equation}
where $p^{min}$ and $p^{max}$ are the boundaries of the parameter
range. The second term in Eq.\ref{llf} increases for larger $p_n^1$
whereas the first term decreases sharply when the goal dynamics
becomes chaotic at $p_n^1 = 3.56$. Therefore the lifetime of the
configuration has a
maximum close to $p_n^1 = 3.56$, the edge of chaos (See
Fig.\ref{lifeb}). This means that if the leader chooses a simple
periodic goal dynamics, it is very likely that he will be successfully
challenged by the follower, since the probability is high that the
goal dynamics of the follower is more complicated than the goal
dynamics of the leader. However, if the goal dynamics of the leader is
highly chaotic, than the follower will also challenge the leader often
since the follower's prediction error is poor. Eq.\ref{llf} is only a rough
estimate since it does not account for the periodic windows
\cite{eajackson} of the logistic map dynamics, which are important at
low noise levels\cite{gmk}.

The passive agent will eventually try to improve his predictions by
switching
on a control; when he does, the environment will follow his
goal only if it is more complicated than the goal of the agent which
is already active. For example if the active agent entrains the
environment to a period four dynamics by using a period four driving
force, the other agent cannot disentrain the environment to a period
two dynamics with a period two driving force, since a period two
entrainment would be disturbed by the period four driving force of the
first agent. However if the second agent tries to entrain the
environment to a period eight dynamics, this may be stable if the
period four driving force of the first agent is taken into account by
the second agent. Another situation arises if the goal dynamics
of the follower is chaotic. It is then difficult to compensate the
periodic driving force of the leader and the environment does not
entrain with the follower during the trial period. 

An overview of
situations in which it is advantageous for the follower to become
active is given in Fig.\ref{lifebb}.  In region $A$ the follower has a more
complicated dynamics than the leader.  Therefore
it can entrain the environment successfully and does not increase its
prediction error by becoming active. In region $B$ the goal dynamics
of the follower is simple compared to that of the leader. The
prediction error increases significantly if the follower becomes
active. In region $C$ the goal dynamics of the follower is chaotic and
usually does not lead to entrainment of the environment. Therefore it
is not advantageous for the follower to switch on its control in
regions $B$ or $C$. In the region $D$ the goal dynamics of the leader
is chaotic. This leads to very large prediction errors for the
follower and makes control for the follower almost always
advantageous.  If the leader chooses a goal dynamics at the edge of
chaos, {\em i.e.} $p_n^1 \approx 3.56$, the probability that the
follower would become active is minimized, since regions $B$ and $C$
contain the entire range of $p_n^2$ values. Therefore the leader
follower-relation is most stable for a goal dynamics at the edge of
chaos.

If we assume that the challenge of the follower is successful if the
complexity of its goal dynamics is larger than that of the leader, the
lifetime of the leader-follower relation increases as $p_n^1$
increases. However as soon as the goal dynamics of the leader becomes
chaotic the prediction error of the follower rises sharply. This
shortens the time span between two trial periods of the follower and
makes it much more likely that the follower challenges the leader
successfully.  Therefore the lifetime of the leader follower
configuration has a maximum for parameters which are close to $p_n^1
\approx 3.56$, the edge of chaos. As may be seen in Fig.\ref{lifeb},
our numerical studies indicate that {\bf leader-follower
configuration} at the edge of chaos is the most stable among all
configurations including those where both agents are active, or both
are passive.

Since a leader follower configuration with a goal dynamics at the edge
of chaos possesses a long lifetime, this configuration may be
considered as an emerging structure. Despite the fact that no social
aspects are included in the system, since both agents only seek to
optimize their own prediction error, a structure with social aspects
emerges, in which one of the agents takes the lead and the other agent
follows the moves of the leader.

\subsection{Two Active Agents}

As noted above, the second agent will not remain passive, no matter
how well he predicts his environment. He will, in time,
switch on his controls in an attempt to improve his prediction powers.  The
system dynamics, for the general case of two active agents,
with different goal dynamics, is complicated.  If both agents use
a control dynamics with a positive exponent, it leads to hyperchaos.  While
it is difficult to estimate
analytically the prediction errors and configuration lifetimes, we
observe in our numerical experiments that the resulting system
dynamics settles into a state in which the environmental dynamics
follows very closely one of the two imposed goal dynamics, in general
the one which is more complicated (see Fig.\ref{competition2}). Thus, if
the goal dynamics of one
agent is a stationary state, while the goal dynamics of the other
agent is a period two dynamics, the environmental dynamics would
follow the period two dynamics. If one agent uses a
highly chaotic goal dynamics, for example $p_n^1 = 4$ and the other
agent a less chaotic goal dynamics, for example $p_n^2 = 3.6$, then
the environment follows the first agent's goal. This ``competition
principle'' would suggest that the agent with the more complicated
goal dynamics has a good chance to outperform his competitor, as
illustrated in Fig.\ref{competition}.  We also observe that the
driving force of the unsuccessful competitor tends to approach a
constant, even if its goal dynamics is not stationary.

In this winner-loser configuration, the prediction error of the
successful competitor is the same as for an active agent without any
competitor, whereas the prediction error of the unsuccessful
competitor is significantly larger. The difference between the two
prediction errors can be estimated by the average difference of the
goal dynamics of the two agents.

If {\bf both} agents are {\bf active}, with one emerging as a successful
competitor, the other as a unsuccessful competitor, a rough estimate of
the prediction error for the successful competitor(winner), labeled here as the
first agent is given by:
\begin{eqnarray}
&&\epsilon_{n,1}^1 = 1 +\exp{(2\lambda^{c,1})} +
 K_n \frac{1}{N_n} + \sum_k
\frac{(\tilde{p}_{k,n}^{model,1})^2}{2\sigma_D^2}\delta_{k,n}\nonumber\\  
&&\mbox{\hphantom{$\epsilon_{n,1}^1=$}}+
\sum_k (\frac{p_{k,n}^{model,1}\sigma_p}{p_n\sigma_D})^2 \frac{N_n}{6}\\
&&\epsilon_{n,m+1}^1 = \epsilon_{n,m}^1 \exp{(2\lambda^{c,1})} +
\epsilon_{n,1}^1\label{epscc2}
\end{eqnarray}
and by
\begin{eqnarray}
\epsilon_{n,m}^2 \approx
\frac{1}{N_n}\sum_{i=n}^{n+N_n}(\frac{x_i^1-x_i^2}{\sigma_D})^2 +
\epsilon_{n,m+1}^1 
\end{eqnarray}
for the unsuccessful competitor(loser).

The prediction error of the winner is the same as for a single active
agent, whereas the prediction error of the loser is generally
dominated by a term which measures the average difference between the
two goal dynamics.  The lifetime $L_{n,m}$ of this configuration can be
estimated from the equation,
\begin{eqnarray}\label{lcc}
L_{n,m} = \frac{c}{\epsilon_{n,m}^2}
\end{eqnarray}

Since the unsuccessful competitor can usually improve his
prediction error by switching his control off, the lifetime of the
winner-loser configuration 
depends only on time span between two trial periods of the
loser, and is determined by the prediction error of the
unsuccessful competitor. The loser may be considered to be
maladapted since his goal dynamics is too simple. In this case the
maladapted agent would lose the competition. Preliminary results
indicate that other types of maladaptation, such as suboptimal
memory or suboptimal complexity, may also lead an agent to lose
the competition.

Our preliminary studies also indicate that time span required
before the winner-loser configuration emerges from a state where both
agents are active increases with the Liapunov exponent of the goal
dynamics of the successful competitor. This observation suggests that an
agent that has a slightly more complicated goal dynamics than his
competitor has the best chance for a rapid improvement of his
prediction error.

\subsection{Adaptation to the Edge of Chaos}

A scenario for transitions between successive configurations is
presented in Fig.\ref{cycle}. It takes the following form:
\begin{itemize}
\item first configuration: {\bf two passive agents}
\item second configuration: One agent switches on its control and becomes
the leader; the prediction error of both agents typically goes down.
One thus arrives at the {\bf leader-follower}
{\bf configuration}.
\item While comparatively stable, the leader-follower configuration
will not persist because as noted earlier, no matter how low the
prediction error of the follower, it will eventually switch on its
control. Most of the time this will not initially increase its
prediction error. However the prediction error of the leader will increase;
the leader will then respond by adjusting his controls, reducing
his prediction error, but increasing that of the ``follower''. The
follower will in turn challenge the leader, who loses
his leadership role. This chain of events is full blown
{\bf competition} which can continue for a long time.  Eventually there
emerges a clear {\bf winner} (the agent with the more complicated goal
dynamics) and a clear {\bf loser}.
\item The {\bf winner-loser} configuration is, of
course not stable, since on the average, the loser, who has the larger
prediction error, will switch off his control, returning the system to
\item the {\bf leader}-{\bf follower} situation in which the
loser significantly improves his prediction error.
\end{itemize}
\ \\

\begin{figure}[h]
\caption[recu15]{
(a): The evolution of the ''goal dynamics'' parameters, as revealed by
numerical experiments with different initial
conditions, denoted by a circle. The agents are seen to adapt
to the edge of the chaotic regime, indicated here by a dashed
line. (b) and (c) are surface plots of the the prediction error of the
two agents as a function of the parameters of the goal dynamics. (d)
and (e) are contour plots of (b) and (c). The parameters are
$\sigma_D=0.006, N_n^1=N_n^2=100, K_n^1=K_n^2=10, p_n =4.0,
c^{control}=0.0002, c^{goal}=10^{-5}$.}
\label{edge}
\end{figure}
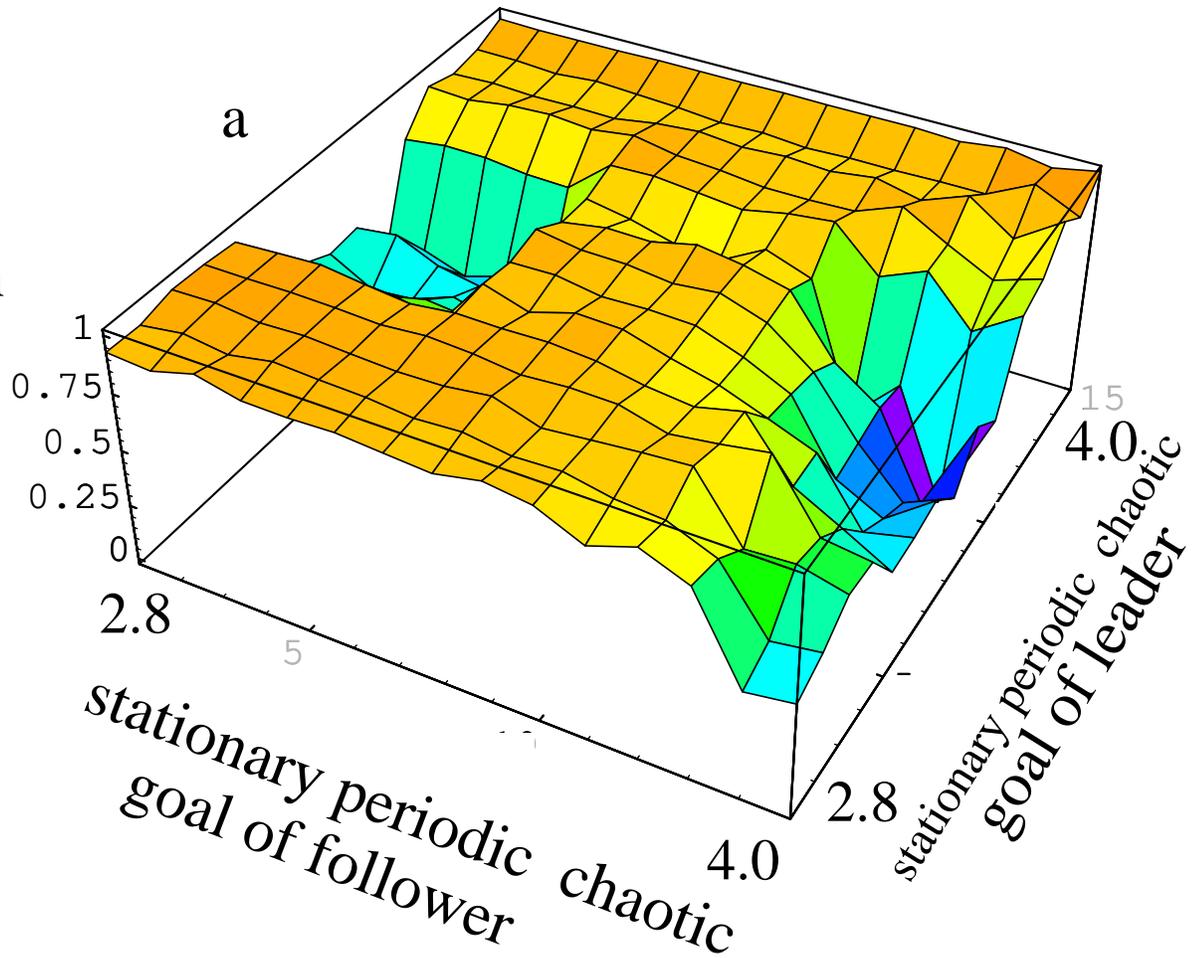
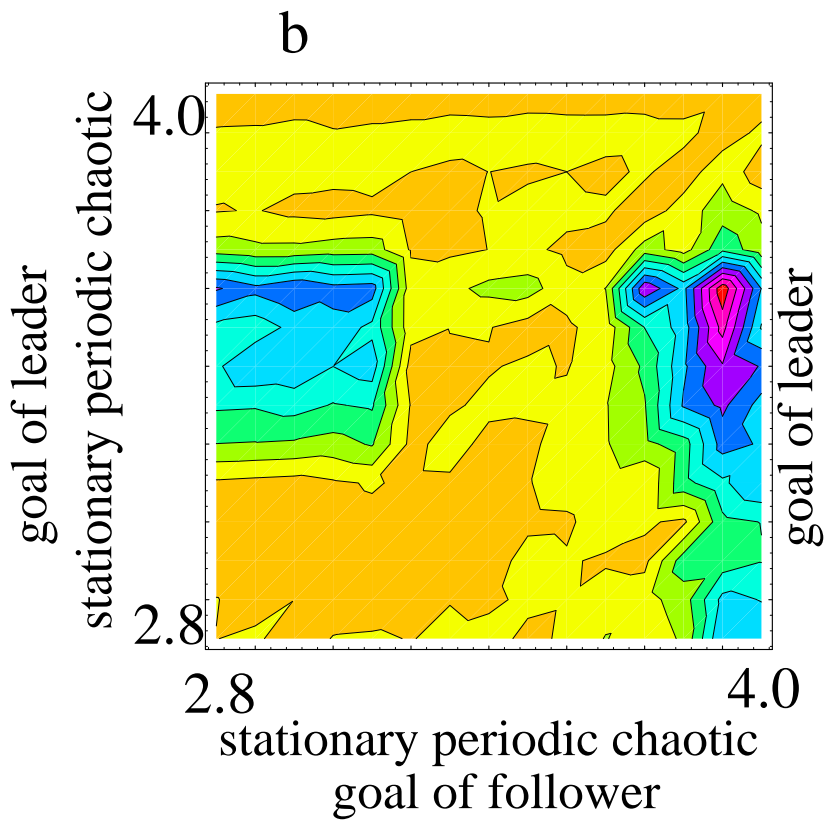
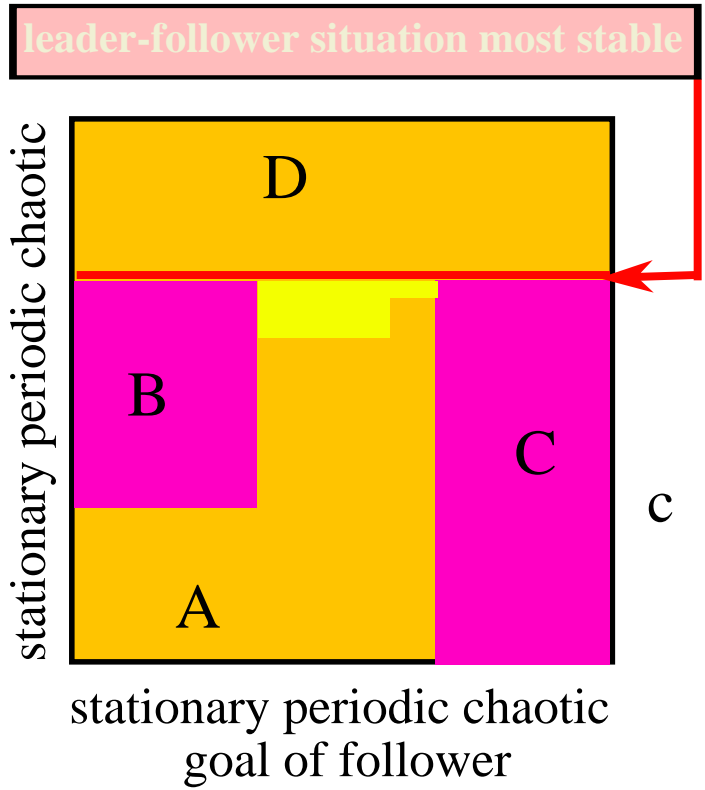
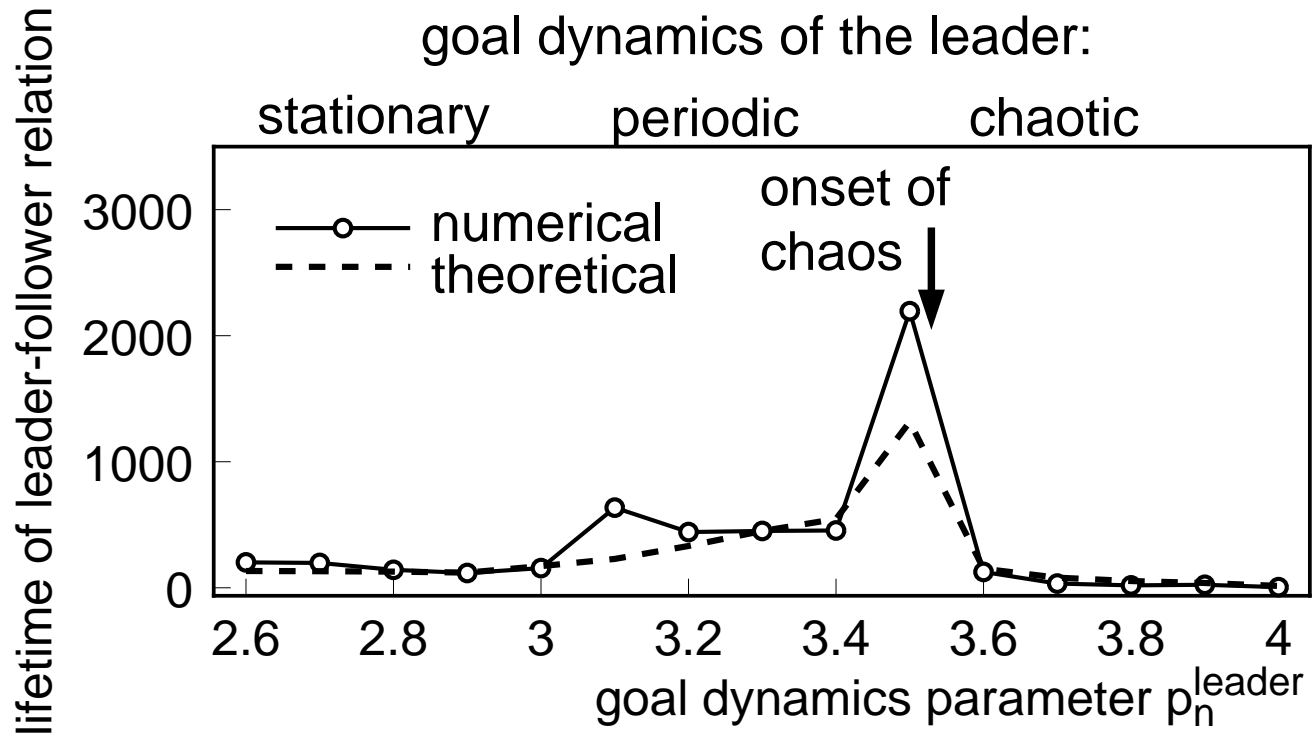
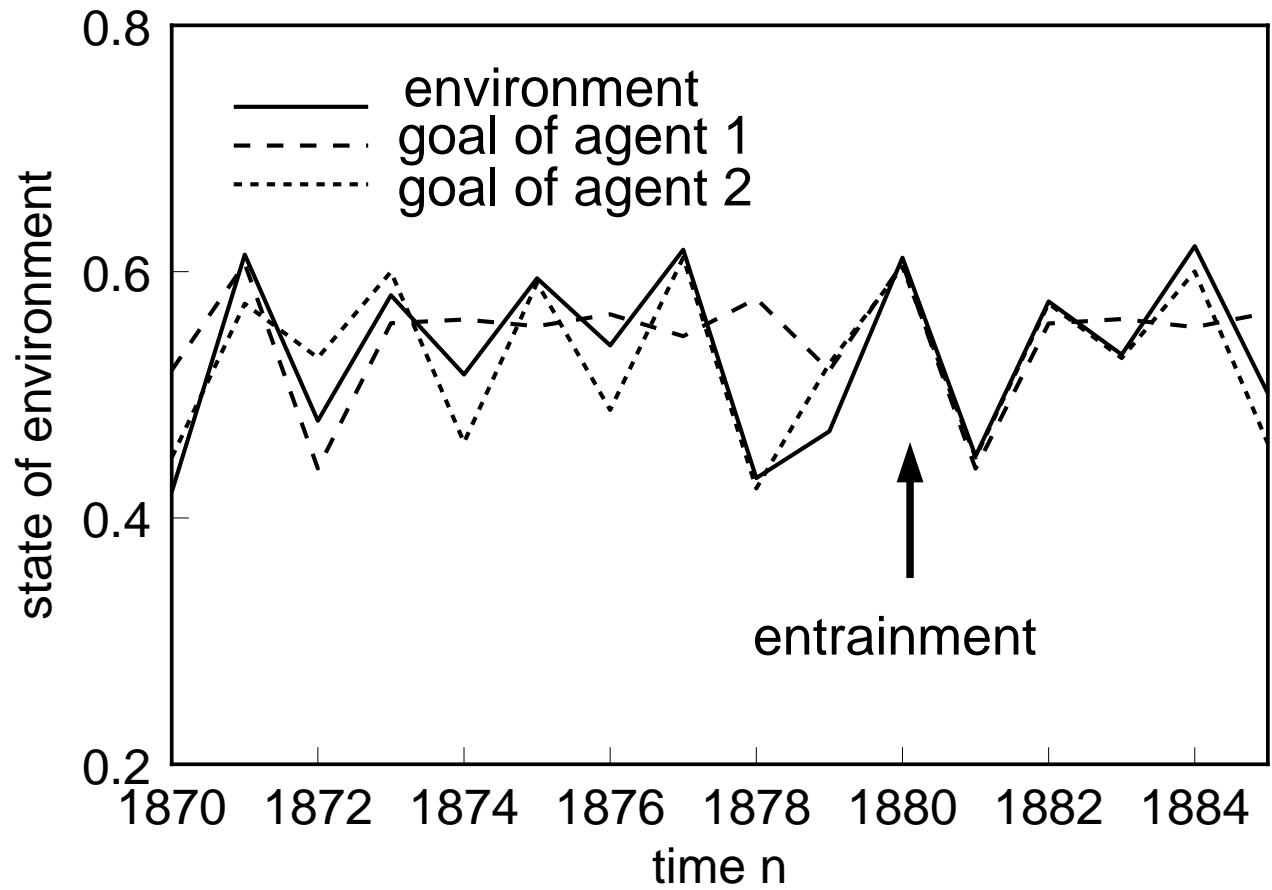
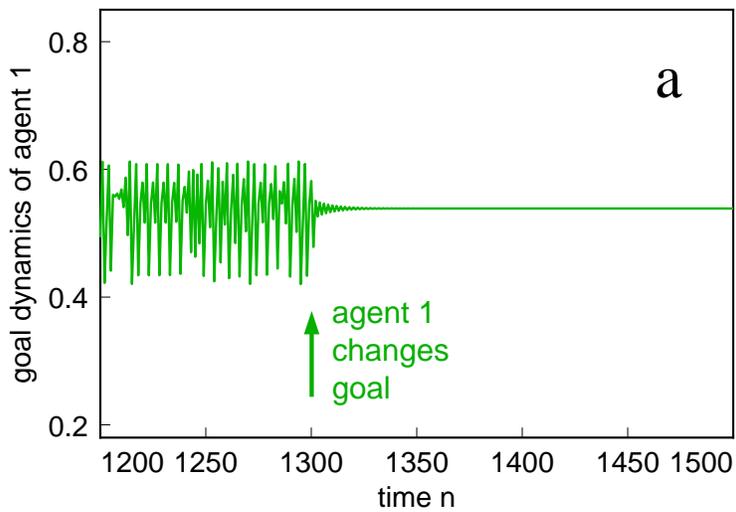
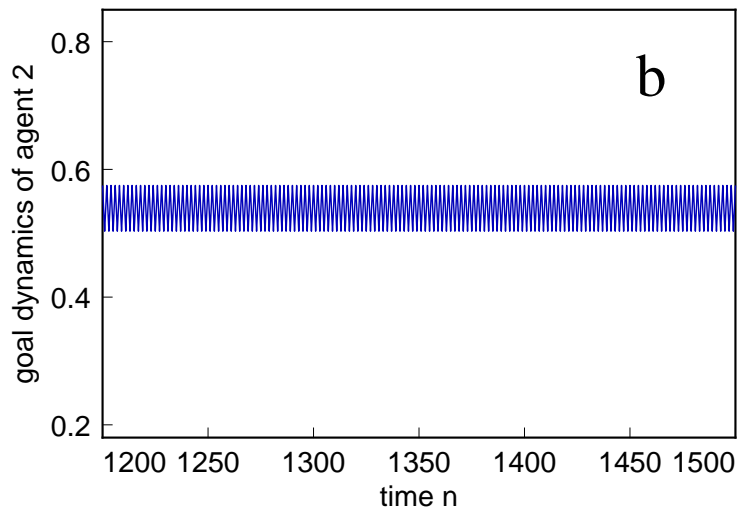
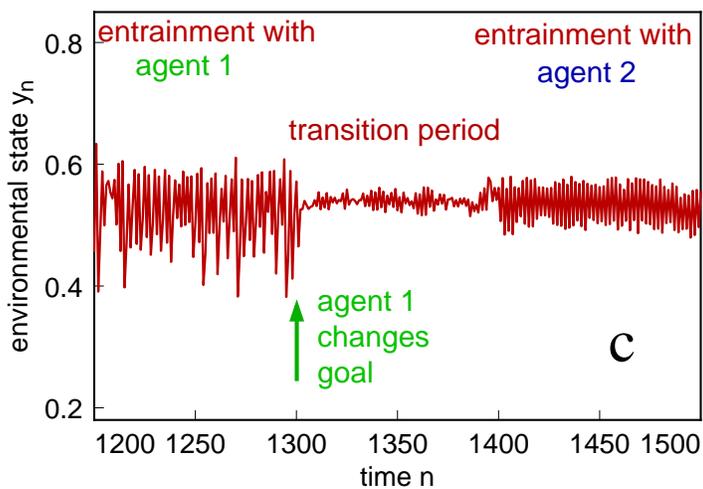
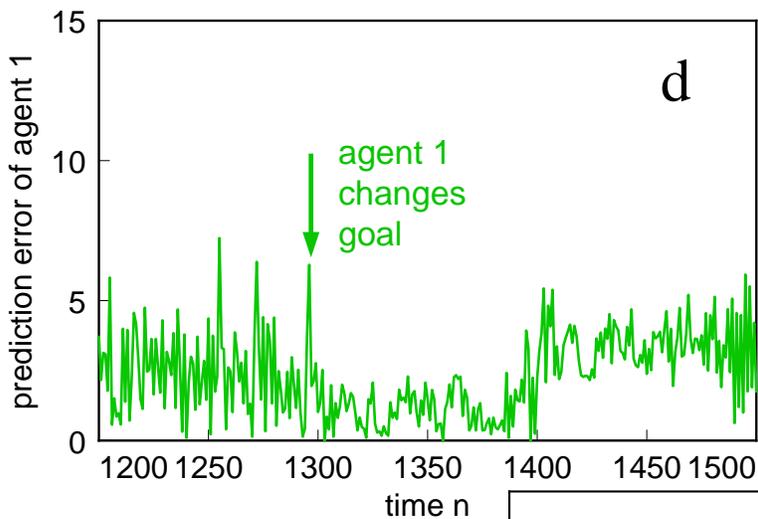
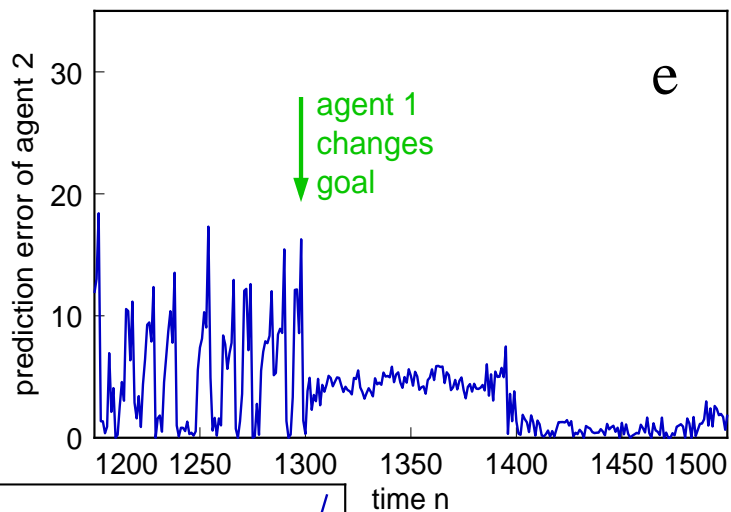
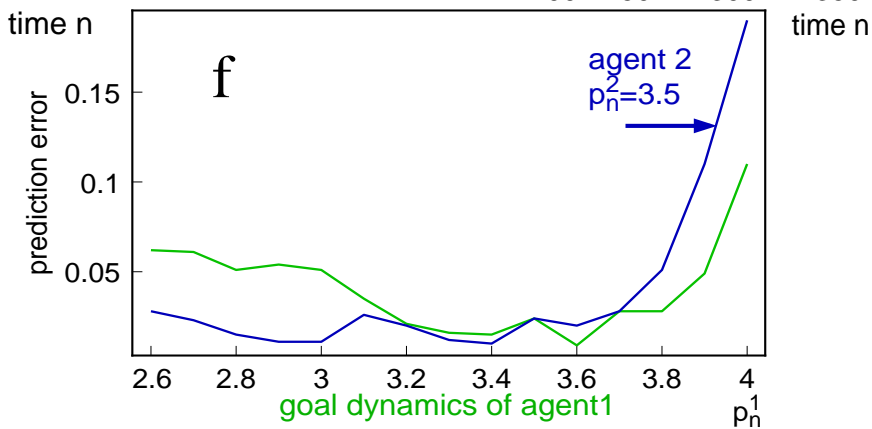
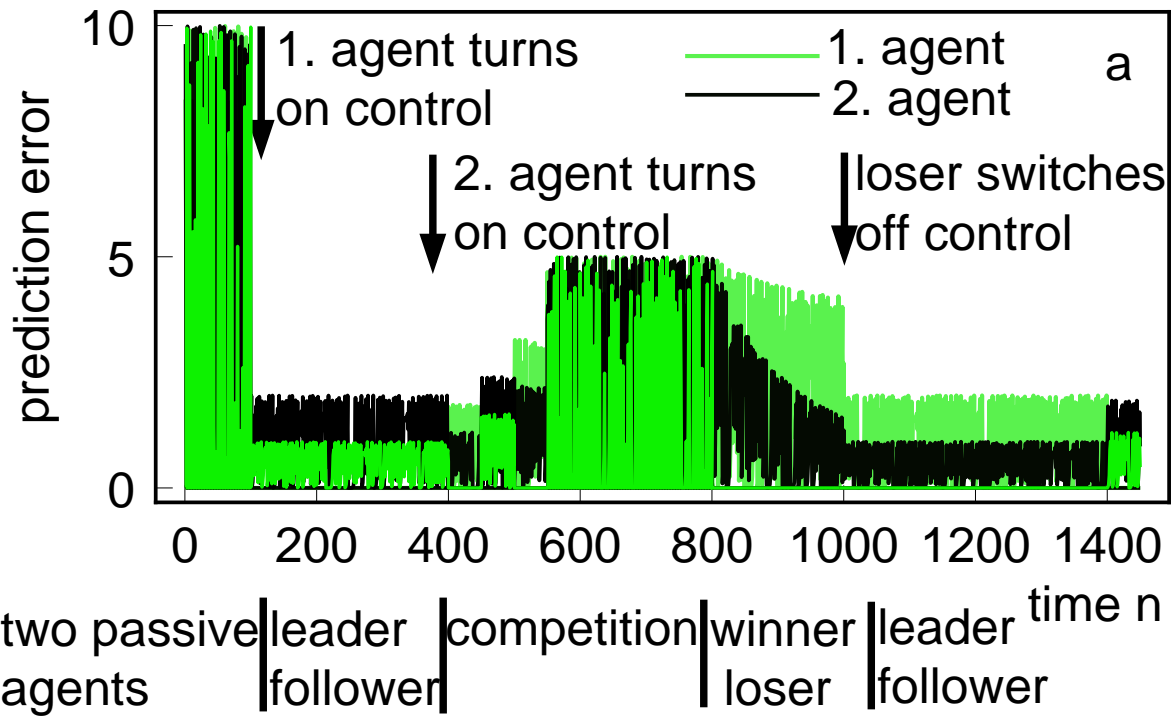
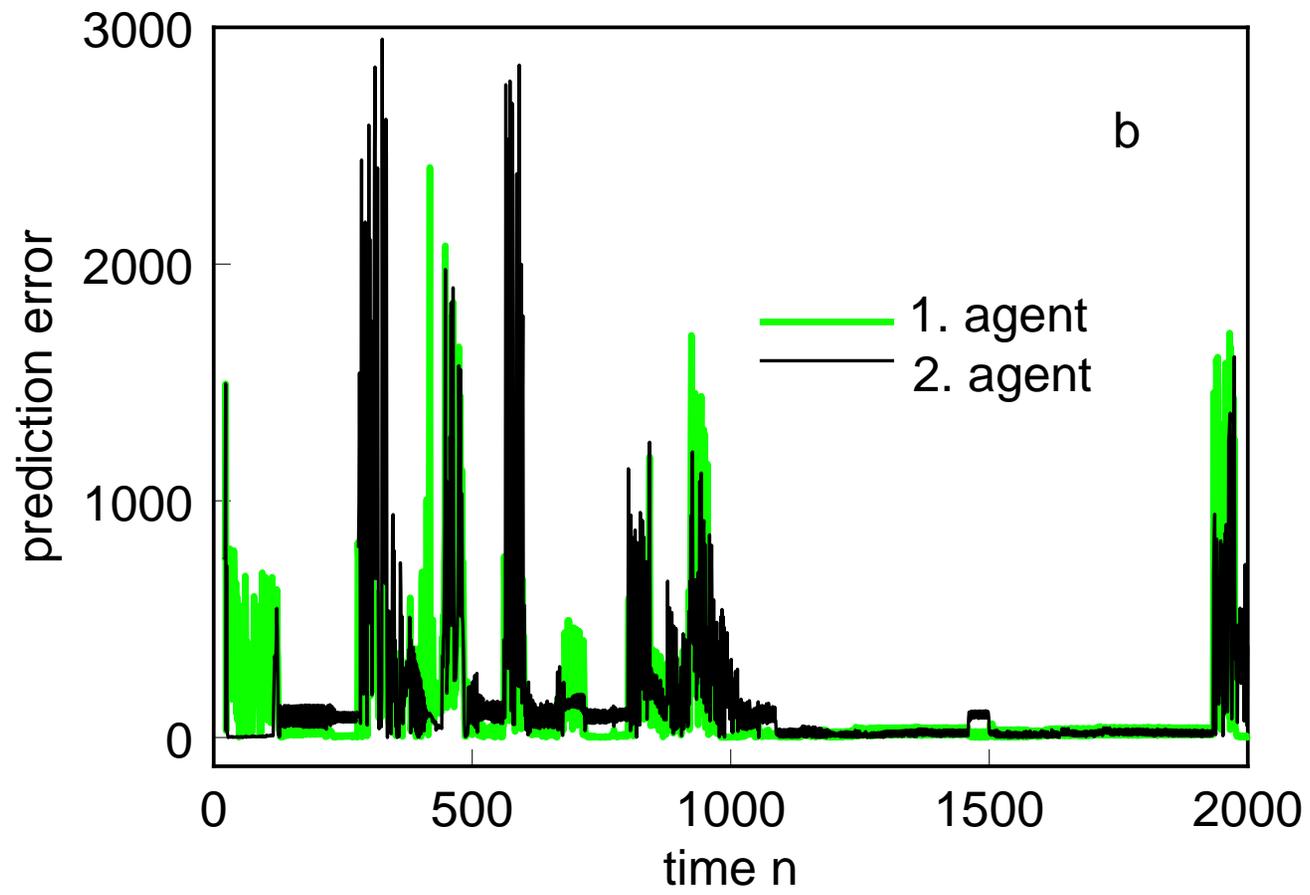
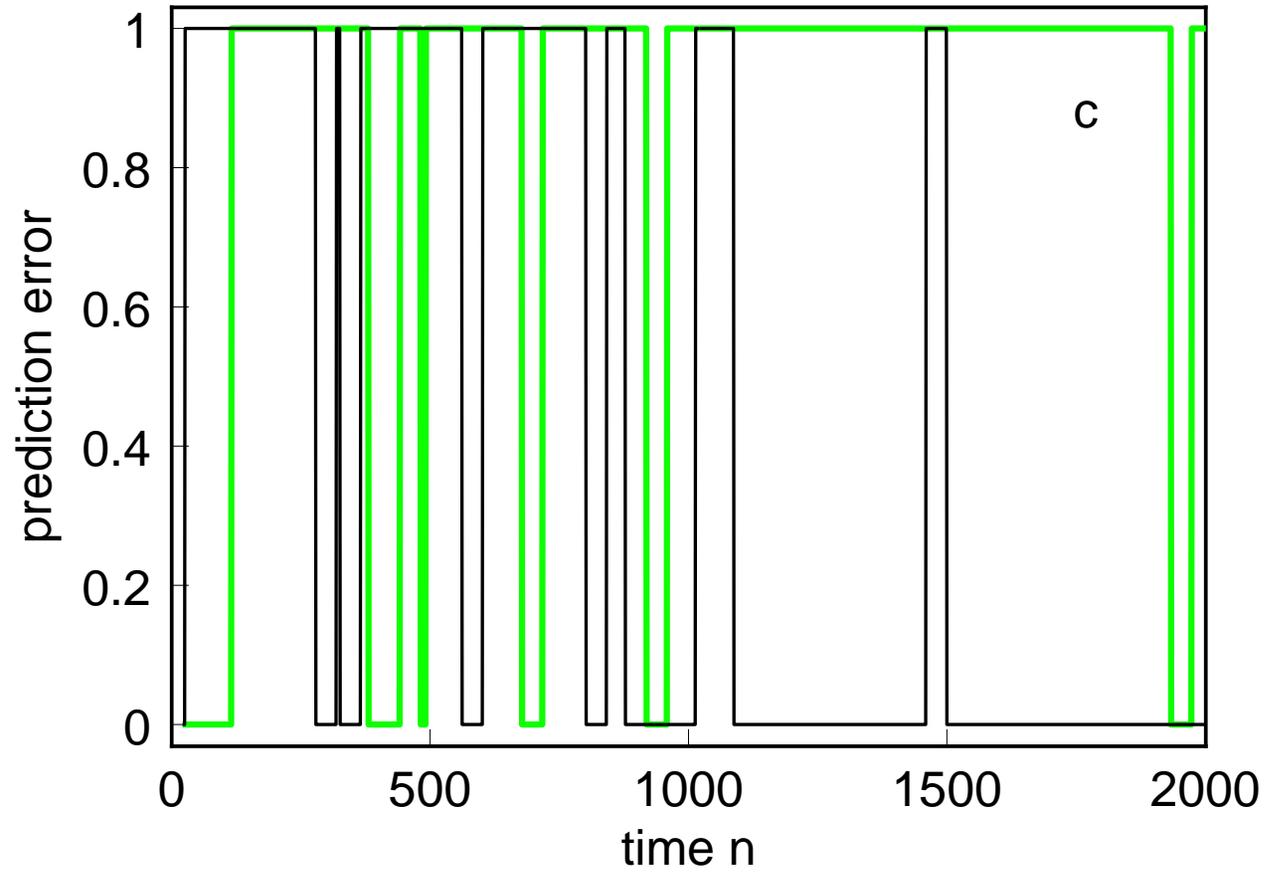
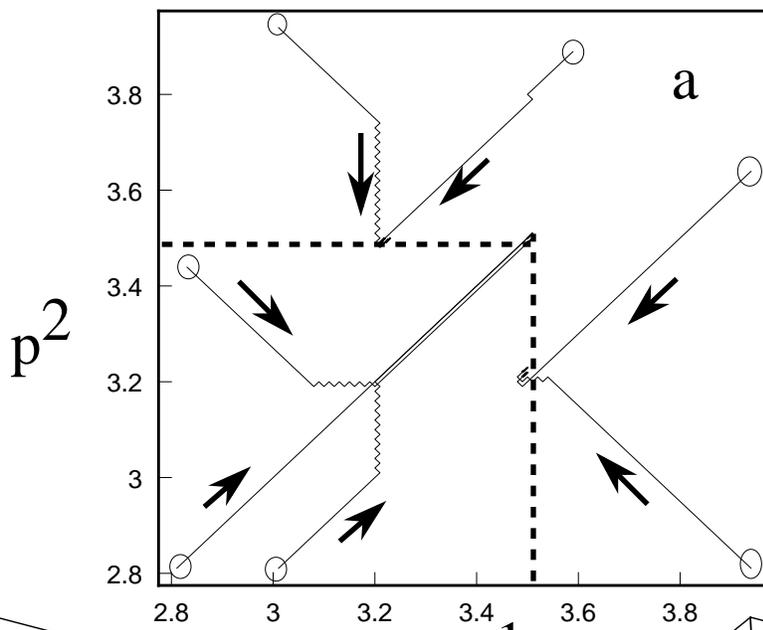
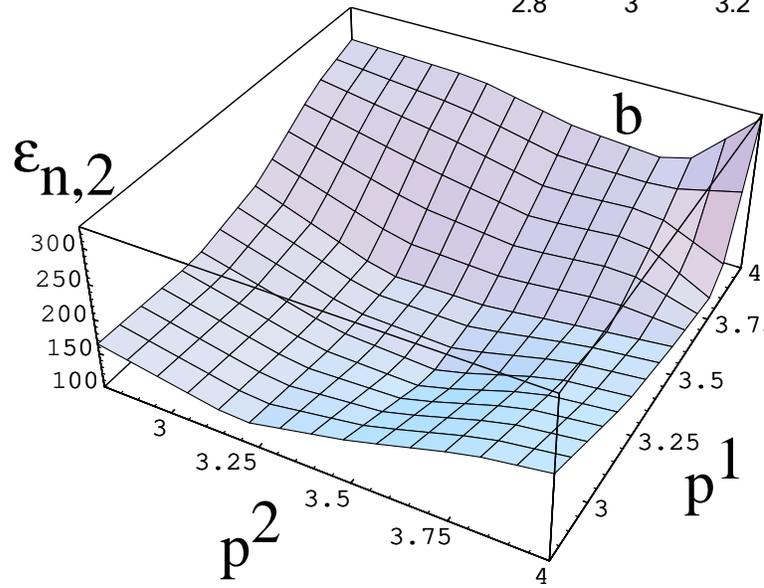
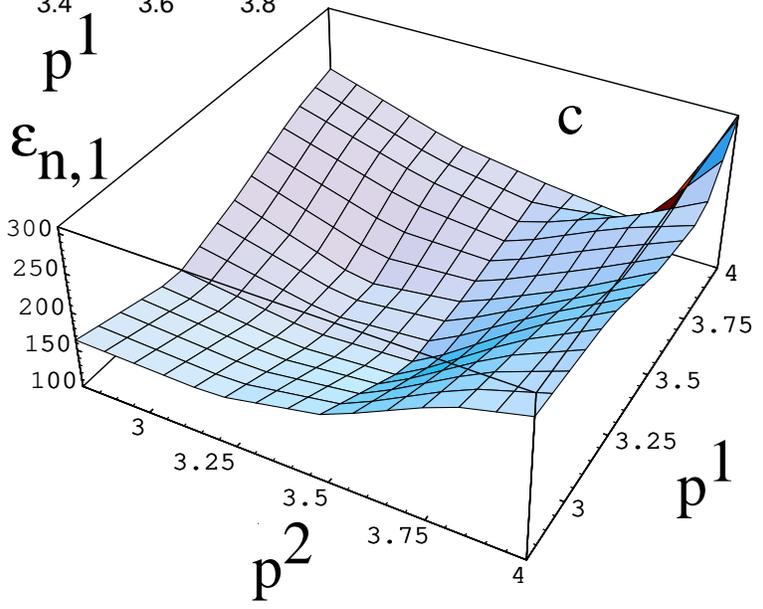
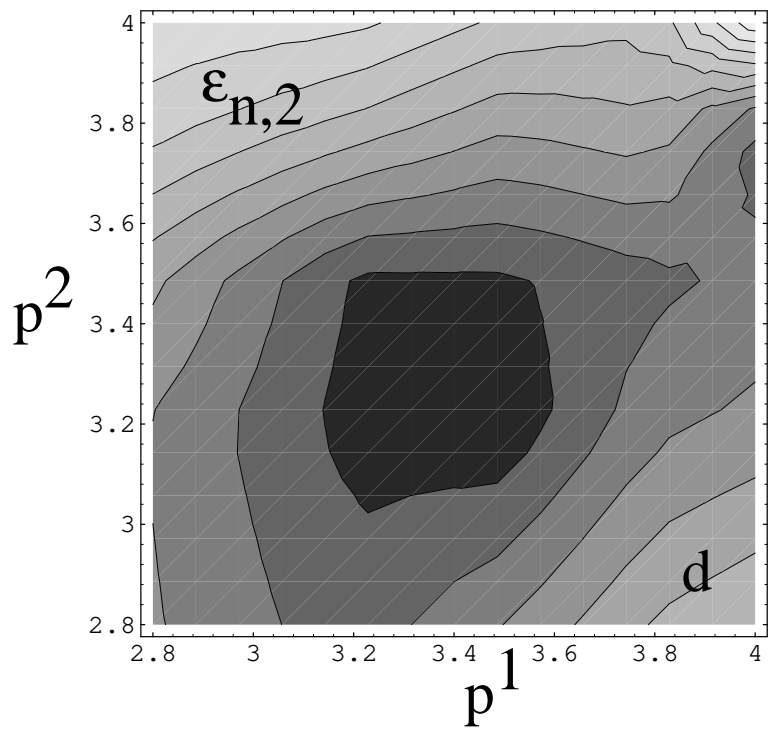
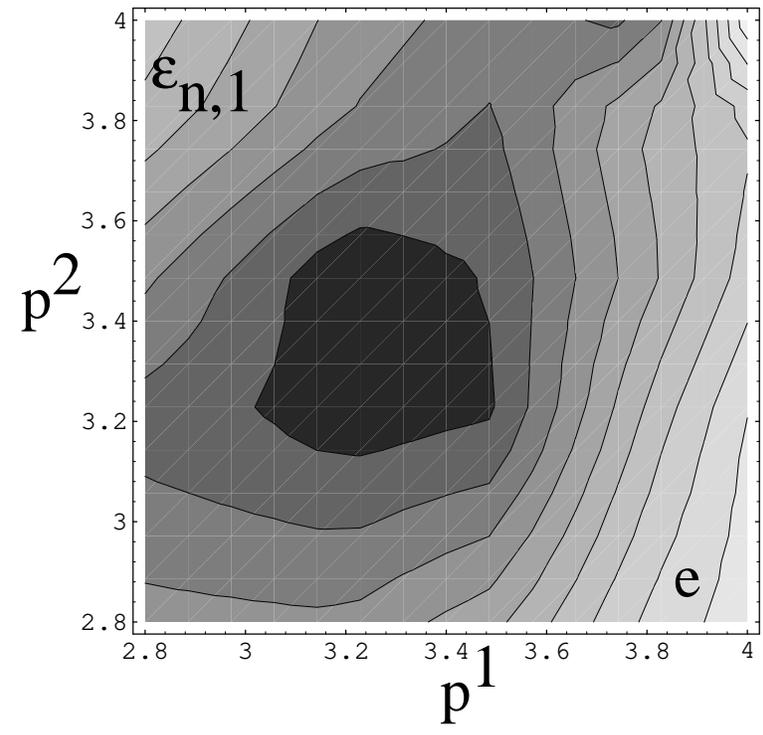

In the course of the competition each agent will find,
by trial and error, that when it improves a goal dynamics with a
complexity greater than that of the apparent leader, its prediction
powers improve as it assumes a leadership role.  On the other hand,
the prediction error of the ``new'' follower increases sharply when
the goal dynamics of the ``new'' leader is chaotic; this condition
will lead it, in turn, to shorten the time between two successive
trial periods, and make it likely that the ``new'' follower becomes
active after a short period of time.
Eventually the leader will use a goal
dynamics at the edge of chaos. This configuration is the most
stable one, since the follower no longer find it advantageous to
switch on his control. This is illustrated in Fig.\ref{edge}.

\subsection{Discovering Other Agents}

It is natural to ask whether one agent can detect the presence of
a second.  If the second is passive, the answer is obviously no.  If,
however, that other agent is active, the answer is yes.  The
strategy for detection is to switch on a control: if the control works,
there is no other active agent
present. If, however, the environment cannot be entrained to a simple goal
dynamics, it is 
likely that another agent is already entraining the environment. Our
adaptive agents use this information in the sense that if they detect
the other agent and therefore experience an increase of their
prediction error during the trial period, they become passive.
However, if they do not detect another agent, {\em i.e.} improve their
prediction error during the trial period, they become active
themselves.

\section{Summary}

The adaptive agents we have studied, although following a
comparatively simple strategy, are able to identify regularities,
generalize and compress observed data by using different sets of
schemata and explore strategies to change their environment.  We find
that quantitative measures of adaptation in a complex environment,
such as complexity, or the learning rate of the agents, approach
limiting values which depend on the rate of change of the environment
and the noise level in the environment. We hope it is possible to use
these measures to test theoretical predictions experimentally in
physical systems, such as phase locked loops, as well as in economic
systems. In our examination of the co-evolution of two agents we
observe the emergence of leader-follower configurations, in which the
leader entrains the environment to a weakly chaotic dynamics.  This
suggests that a primary goal, such as a small prediction error, or a
large return for an investment in an airline company in a highly
competitive market, may cause secondary goals which are chaotic: the
chaotic goal dynamics for the environment in this study or a chaotic
dynamics of the pricing of products such as air fares.

It will be interesting to explore the relationship of the tendency of
the agents to move to the edge of chaos with the general idea of
adaptation to the edge of chaos as discussed by
Langton\cite{edgelangton}, Packard\cite{edgepackard},
Kauffman\cite{edgekauffman}, and Crutchfield\cite{crutchfield}.

It is also interesting to speculate on the applicability to real world
situations of some of the results we have obtained from our toy
model. For example, we have seen that active adaptation, using
control to improve prediction, is under most circumstances the
preferred strategy. This finding seems in accord with experience,
whether one is analyzing the way an infant develops predictive
power by exercising control of its immediate environment
(parents) through cries or smiles, or analyzing the behavior of two
interacting adults. Consider, too, the attempt by traders on a stock
market in the early minutes after the opening, to exercise control
and improve their short term predictive powers, by carrying out a
series of trades designed to probe, actively, likely subsequent
trading patterns on that day.

To cite another example; we have seen that in active competition
between agents, the agent with the more complicated strategy will win.
This finding accords with experience on the political scene. It helps
suggest why, for example, the Serbs have proved so successful to date
in pursuing their strategy of ethnic cleansing; their strategy may be
regarded as one involving a series of controlled experiments which
enable them to predict the UN response.

Although the system which we study is simple, it makes possible a
quantitative comparison between numerical and analytical results.  We
hope that some of our findings for simple adaptive agents in a chaotic
environment are applicable to the behavior of real adaptive agents in
complicated economic and/or biochemical systems.  Of course, to extend
our approach to economic systems it is important to incorporate into
our model the cost of constructing a model, exercising control, etc.
Still, even at the present level, our approach would seem to provide
insight into the success of technical trading systems. Technical
trading systems usually use time bars to describe the spread of values
of a time series, whereas in physics and engineering the variance is
commonly used for that purpose. It can be shown\cite{adaptation} that for
chaotic time series time bars are maximum likelihood estimates of the
spreading of the data in contrast to the variance. To the extent that
economic time series are chaotic, this could explain why it is
advantageous to use time bars for analyzing their behavior.

We intend in the future to extend in a number of ways the numerical
experiments presented here. We plan to study the competition between
agents of markedly different adaptive capacities ( as manifested both
in the ability to model the environment and to control it) and to
extend our approach to many interacting agents in order to examine
possible collective behavior. While we have seen that the outcome of
the present simulations of the interaction between two agents appears
to lead to either a ``win-win'' or a ``win-lose'' situation, we
anticipate that the actions of a powerful maladaptive agent can
lead to a ``lose-lose'' situation, and it will be interesting to
specify the conditions under which this comes about. As our program is
currently written, no matter how well an agent is doing, it will, over
time, seek to improve its predictive powers by changing its strategy,
which means that all the configurations we have considered are
metastable. We therefore plan to modify our innovation paradigm,
Eq.(5) by introducing a threshold for change; we expect that the
resulting ``happy agent'' configuration may lead, in some
circumstances, to stable ``win-win'' configurations. Finally, in order
to make more direct contact with economics, we intend to introduce
both a cost of computation and a cost of control in our numerical
experiments.

In another direction, we call the attention of the reader to a closely
related set of independent numerical experiments carried out by Kaneko
and Suzuki\cite{kaneko} on a model for the evolution of the complex
song of a bird. They use a simple logistic map for the song dynamics
and consider a two person game between competing ``birds''. Their
agent ``birds'' adapt to one another, exercising control through their
songs. Kaneko and Suzuki find that the dynamics of the complex song
evolves toward the edge of chaos. It will be interesting to explore
the relationship between their simulations and our own, and to see to
what extent a complicated environment might play a role in that
evolution.

We wish to thank John Miller for his helpful advice in the early stages of
our specification of this model as well as his cautions
concerning its immediate applicability to economics, and Gottfried
Mayer-Kress for advice on the development of our graphic displays
and for stimulating discussions on these and related topics.  We thank Bill
Fulkerson for a critical reading of a preliminary version of this
manuscript, and a number of helpful suggestions. This work was
begun at the Santa Fe Institute with support from a Robert
Maxwell Professorship, and has been subsequently supported both by the
Santa Fe Institute and by the Center for Complex Systems
Research at the Beckman Institute of the University of Illinois at Urbana
Champaign;
we thank both institutions for their support. The present version of
our manuscript has profited from the informal remarks of our fellow
participants at the Integrative Themes Workshop of the Santa Fe Institute, whom we thank for
their advice and encouragement.

\clearpage
\section{Figure Captions}

\begin{figure}[h]
\caption[recu1]{A typical time series of the environment specified by
Eq.\ref{envnoise1}, with $\sigma_D=.006$, $\sigma_p=.0006$, and
$2<p_n<4$}
\label{fignoise} 
\end{figure}

\begin{figure}[h]
\caption[recu2]{(a):
A state space representation, $y_{n+1}$ versus $y_{n}$, of the time
series plotted in Fig.1. (b): The dashed line represents the exact map
$f(y_n)$, the dotted line is a linear interpolation between the
observed events, $f_{j,{n}}^{grid,agent}$, and the continuous line is a
Fourier approximation of the dotted line, $f^{model,agent}_{{n}}(y_{n})$.
The observed events are in the region
$0.2<y_n<.92$. The extrapolation of the model outside this region is
not smooth, but can be made smooth with an appropriate choice of
$d_{0,n}^{model}$ and $d_{0,n}^{model}$; (c): an enlargement of the
center region of (b); (d): The predicted evolution of the time series
shown in Fig.1. The dot-dashed line is the system behavior, while the
triangles denote the predicted values; the errors associated with the
prediction are also shown.}
\label{interpolation}
\end{figure}

\begin{figure}[h]
\caption[recu3]{(a): The single step prediction error as a function of
the number of model parameters $K_n$; all Fourier coefficients with
$k<K_n$ are included in the model. The parameter values are
$\sigma_D=0.006, \sigma_p=0, N_n=100, p_n=4$. (b) The size of the model
Fourier coefficients and the size of their standard deviation as
a function of the model parameter $k$. (c): The prediction error as a
function of the number of events used to extract the model. The
parameter values are $\sigma_D=0.006, \sigma_p=0.0003, K_n=10$.  The
dashed lines represent numerical results, and the continuous line is
the analytic result.}
\label{optimal}
\end{figure}

\begin{figure}[h]
\caption[recu4]{Experimental detection of adaptive agents. At time 1100
the parameter of the environmental dynamics is suddenly changed from
4.0 to 3.5. The single step prediction error increases significantly,
but returns to its original value within 100 steps, which equals
$N_n$. The parameters are $\sigma_D=0.006, \sigma_p=0, N_n=100, p_n=4,
K_n=10$}
\label{recovery}
\end{figure}

\begin{figure}[h]
\caption[recu5]{Multiple step prediction error of a
passive agent as a function of the number of time steps and the
Liapunov exponent of the environmental dynamics.}
\label{sp} 
\end{figure}

\begin{figure}[h]
\caption[recu6]{Situations in which it is
advantageous for an agent to become active. Depicted is the the 5-step
prediction error as a function of the parameter of the goal dynamics
and the parameter of the environment for the case without control (a)
and the case with control (b) and (c), the latter being an enlargement of
(b); (d) a schematic illustration of the different
regions.  The parameter values are $\sigma_D=0.006,
\sigma_p=0, N_n=100, K=10$.}
\label{sc} 
\end{figure}

\begin{figure}[h]
\caption[recu7]{Numerical simulations of active adaptation.  (a) The
evolution of the environment, $y_n$, subject to an active agent
with a goal dynamics $p_n^{goal}=2.6$, illustrated in (b), which turns on
its control at $t = 1100$, (c) depicts the resulting
prediction error, which is seen to drop significantly soon after the
environmental dynamics becomes entrained by the goal dynamics.
(d)-(f) show a similar series of plots for the case that the goal dynamics
is chaotic ($p_n^{goal}=3.8$). The parameter values are
$\sigma_D=0.006, \sigma_p=0, N_n=100, p_n=4, K=10$.}
\label{control} 
\end{figure}

\begin{figure}[h]
\caption[recu8]{(a): The single step prediction error of an
active agent as a function of the parameter, $p_n^{goal}$ of the goal
dynamics, for
$\sigma_D=0.0002, \sigma_p=0., p=4., N_n=100, k_n=10$ (continuous line
denotes the numerical results, dashed line, the statistical estimate). (b)
The
exact map (dotted line), is compared to the linear interpolation (dashed
line), and
the model map (continuous line) versus $y_n$. (c): An enlargement of
(b).}
\label{withcontrol} 
\end{figure}

\begin{figure}[h]
\caption[recu9]{The prediction error
of the leader and follower in the leader-follower configuration, as a
function of
the goal dynamics of the leader. The circles represent the results of
our numerical experiments; the continuous line the analytic result.
The parameter values are $\sigma_D=0.006, N_n^1=N_n^2=100,
K_n^1=K_n^2=10, p_n =4.0$}

\label{predlf} 
\end{figure}

\begin{figure}[h]
\caption[recu10]{An overview of situations in which it is
advantageous for the follower to become active. Depicted is the ratio
between the 5-step prediction error in the trial period of an active
follower and mean prediction error of a passive follower as a function
of the parameter of the goal dynamics of leader and follower The
surface plot (a) and the contour plot (b) are numerical results. (c)
illustrates the different regions schematically.  The parameters used
in the simulation are $\sigma_D=.006, p_n=4, K_n^1=K_n^2=10,
N_n^1=N_n^2=100$.}
\label{lifebb} 
\end{figure}

\begin{figure}[h]
\caption[recu11]{The lifetime of the leader
follower configuration as a function of the parameter of the goal
dynamics of the leader, when the parameter of the goal dynamics of
the follower is randomly chosen in the interval $2.6 < p_n^{c,1} < 4$.
The continuous line represents the numerical result, the dashed line, the
analytic result. The parameters are $\sigma_D=.006, p_n=4,
K_n^1=K_n^2=10, N_n^1=N_n^2=100$}
\label{lifeb} 
\end{figure}

\begin{figure}[h]
\caption[recu12]{
Numerical simulation of an environment subject to active agents which
possess similar goal
dynamics. Most of the time, the environment is disentrained; however it
locks with
the two agents when they utilize similar goal dynamics, such as at time
$n=1880$. The parameters are $\sigma_D=0.006, N_n^1=N_n^2=100,
K_n^1=K_n^2=10, p_n =4.0, p_n^1=p_n^2=3.8$}
\label{competition2}
\end{figure}

\begin{figure}[h]
\caption[recu13]{
Two typical situations involving active agents. One agent has a chaotic
goal dynamics up to time $n=1300$ and
then switches to a stationary goal(a), whereas the other agent has a
periodic goal dynamics (b) for the whole time span. (c) shows the
response of the environment. Up to time $n=1300$ it is entrained by
the chaotic dynamics of the first agent. The first agent is the winner.  
When this agent switches to a stationary goal, the environment becomes
entrained by the other agent after a transition period of 100 steps,
which equals the memory of the agent. (d) and (e) show the 5-step
prediction error of the agents.(f):This plot shows the 5-step
prediction error of both active agents as a function of $p_n^1$, for a
fixed
goal dynamics $p_n^2=3.5$ of one agent.  We see that the agent with the
larger $p_n$ value, {\em i.e.} the
more complex goal dynamics, is the winner. The parameters are
$\sigma_D=0.006, N_n^1=N_n^2=100, K_n^1=K_n^2=10, p_n =4.0$}
\label{competition}
\end{figure} 

\begin{figure}[h]
\caption[recu14]{
(a): This schematic figure shows a typical sequence of events (b): the
results of a numerical simulation for the parameter values
$\sigma_D=0.006, N_n^1=N_n^2=100, K_n^1=K_n^2=10, p_n =4.0, p^1=3.5,
p^2=2.8, p^{control}=0.0002$ (c): the dynamics of the control
coefficients which indicate whether the agents are active or passive.}
\label{cycle}
\end{figure}

\end{document}